\documentstyle[12pt]{article}
\begin{document}
\begin{center}
{\bf {\Large
On flux phase and N\'eel antiferromagnetism
in the {\em t}-{\em J}-model}}
\\[0.9cm]
I.S.~Sandalov,\\[0.3cm]
{\it
Department of Physics, Uppsala University, Box 530, S-75121 Uppsala,
Sweden \\
and\\
Kirensky Institute of Physics of RAS, 660036 Krasnoyarsk, Russia}\\[0.5cm]
and M.~Richter$^*$,\\[0.3cm]
{\it
Department of Physics, Uppsala University, Box 530, S-75121 Uppsala,
Sweden}\\[1.5cm]
\end{center}
{\small $^*$present address: MPG Research Group ``Electron Systems'',
Physics Department, University of Technology, D-O-8027 Dresden, Germany }
\\[3.0cm]
\begin{tabular}{ll}
{\bf running title:~} & $t$-$J$-model: flux phase, antiferromagnetic
\\[1.0cm]
{\bf keywords:~}      & $t$-$J$-model \\
                      & antiferromagnetic order \\
                      & flux phase \\
                      & phase diagram \\[1.0cm]
{\bf PACS:~}          & 7420\\
		      & 7510L \\

\end{tabular}

\newpage
\subsubsection*{Abstract}
We reanalyse the mathematical formulation of the flux-state problem  within the
$t$-$J$ model.
The analysis of different parametrizations in the functional representation
shows that (i)  calculations which take
 into account constraints for the number of on-site-available states are
 describing quasiparticles in terms of wrong
local statistics, and contain gauge non-invariant objects;
(ii) application of the projection technique in the slave-boson(fermion)
representation reproduces the correct statistics, and is exactly equivalent
to the  conventional diagram technique for Hubbard operators;
(iii) it is necessary to introduce an additional equation for the effective
hopping amplitude  for the flux phase.

  With the technique for Hubbard operators, which allows one to separate
  charge and spin channels
we construct the mean-field equations for a flux-like
state  which coexists with N\'eel antiferromagnetism (AF).
  The formal    analysis shows that
  the equations for real and imaginary parts of the effective hopping
  amplitude  are inconsistent for  any $\theta \neq 0$
  (including $\theta =\pi /4$ which gives flux 1/2).
  The hopping amplitude is slightly supressed by exchange renormalization.
  The N\'eel magnetization $m$ decreases with increasing concentration
  of holes.
  The region where antiferromagnetism exists is decreasing
  with increasing bare hopping parameter. For $t/J \geq (t/J)_{c}$,
  there are no
  magnetic solutions {\em below} a certain concentration of holes.
Moreover, at large values of $t/J$ ($t/J \geq 1.2$) the region,
where antiferromagnetic order can exist, is found to be very
narrow and situated near the critical concentration of holes
$\delta=1/3$.
Since it is known that in experiment a very small amount of holes
(about 2\%) destroys antiferromagnetism, we conclude,
that the $t-J$-model is not good enough for the
description of copper oxides.

\newpage
\subsubsection*{1. Introduction}
 High-$T_C$ superconductors seem to be good examples of systems
with strongly correlated electrons (see, for example, the discussion in$^{1}$).
The $t$-$J$ model has been  proposed to describe the properties of holes
in the CuO$_2$ planes of these materials$^{2,3}$.
A  cluster derivation of this effective Hamiltonian$^{4}$ shows that
the $t-J$ model,
\[
H=H_t+H_J~,
\]
\begin{equation}
H_t=-t \sum_{\langle{}ij\rangle{}}
(1-n_{i,-\sigma})c^+_{i\sigma}c_{j\sigma}(1-n_{j,-\sigma})~,
\end{equation}
\[
H_J=2J \sum_{\langle{}ij\rangle{}} ({\bf s}_i {\bf s}_j - \frac{1}{4}n_in_j)~,
\]
can describe the real situation only if the direct exchange interaction between
$d$ and $p$ orbitals is sufficiently large and of antiferromagnetic sign.
The generalized Hubbard model$^{3}$ is very complicated, but the model (1)
contains the most important physical features of highly correlated electron
systems.
The main feature of (1) is that the hopping probability for electrons
strongly depends on the type of magnetic ordering of the electrons.

The Hubbard model with single-site repulsion
$U\rightarrow \infty$ $(J = 0)$ in the ferromagnetic phase
 has been considered by Batyev$^{5}$. He
represented  the hopping term in (1) in the form
\[
H_t=-t\sum_{\langle{}ij\rangle{}} \Psi^+_i \Psi_j
[(\frac{1}{2}+s^z_i)(\frac{1}{2}+s^z_j)+s^+_is^-_j]
\]
with the constraint
\[
\hat{N}_2=\sum_i \Psi^+_i \Psi_i (\frac{1}{2}-s^z_i)=0~,
\]
forbidding doubly occupied states.
Assuming a state with a magnetization varying slowly from point to
point, he considered the following problem: How does the spin affect the hole
motion?
Using a probe wave function of the form
\begin{equation}
\Phi=\prod_i \left\{ \left( \begin{array}{c} u_i \\ v_i \end{array}
             \right) (1-\Psi^+_i \Psi_i) +
             \left( \begin{array}{c} 1 \\ 0 \end{array} \right)_i
             \Psi^+_i \Psi_i \right\} \Phi_0~,
{}~~~|u_i|^2+|v_i|^2=1~,
\end{equation}
where the state $\Phi_0$ depends only on the number of holes, he has
obtained the effective Hamiltonian $H_{{\it eff}}$ for holes
\[
(\Phi,H\Phi)=(\Phi_0,H_{{\it eff}}\Phi_0)~,
\]
\begin{equation}
H_{{\it eff}}=-\sum_{\langle{}ij\rangle{}}
\Psi^+_i \Psi_j (u_iu^*_j+v_iv^*_j)~.
\end{equation}
Expanding $v_i-v_j$ and $u_i-u_j$ in powers of ${\bf R}_i-{\bf R}_j$,
Batyev found
\begin{equation}
H_{{\it eff}}=\frac{1}{N}\sum_{\vec{p}}
 \left[\frac{1}{2}({\bf p}-{\bf A})^2+U({\bf r})\right]
\end{equation}
in ${\bf p}$-representation, where
\begin{equation}
U({\bf r})=\frac{1}{2}[(\nabla \Phi^*, \nabla \Phi)-\vec{A}^2]~,~~~
\Phi=\left( \begin{array}{c} u\\v \end{array} \right)~,
\end{equation}
\begin{equation}
{\bf A}({\bf r})= -\frac{i}{2}[(\Phi^*,\nabla \Phi)-(\nabla \Phi^*, \Phi)].~
\end{equation}
As seen here the values and directions of ``electric''
and ``magnetic'' fields are determined by spin wave functions.

A more general formulation  starts from the representation of the
partition function as
a functional integral over coherent states$^{6}$. To clarify the
difficulties in the application of the functional approach to systems
with strongly-correlated electrons  we briefly outline this approach.
The basic idea is as follows$^{7}$:
Any state of the system can be represented in the form
\[ |\Psi_i\rangle{}=T_i|0\rangle{}~, \]
where $|0\rangle{}$ is the highest vector of the representation of the
corresponding group.
$T$ contains an expansion over the set of generators of the
algebra under consideration (or superalgebra $\equiv$ graduated
algebra).
The projector
\begin{equation}
 P(\alpha)=T(\alpha)|0\rangle{}\langle{}0|T^+(\alpha)~,
{}~~~P^2=P
\end{equation}
is introduced, with the requirement
\begin{equation}
\int d\mu(\alpha) \, P(\alpha) =\hat{1}~.
\end{equation}
Equation (8) is a requirement for the form of the
measure $d\mu(\alpha)$ over the group parameters.
Any group operator can be represented in terms of parameters
$\alpha$ entering $T(\alpha)$.
The partition function is then written as a product$^{6}$
\[
\begin{array}{lr}
Z={\rm Sp}\, e^{-\beta H} = &
  {\displaystyle \int \prod_i}
  d\mu_{\tau_i} \, \langle{}0|T^+_{\tau_N}e^{-(\tau_1-0)H}
T_{\tau_1}|0\rangle{}
                \langle{}0|T^+_{\tau_1}e^{-(\tau_2-\tau_1)H} T_{\tau_2}
                |0\rangle{} \ldots\\[0.3cm]
&              \ldots \langle{}0|T^+_{\tau_{N-1}}
               e^{-(\tau_N-\tau_{N-1})H} T_{\tau_N}|0\rangle{}~,
\end{array}
\]
where $\tau_{i+1}-\tau_i=\beta/N\equiv\Delta\tau$ ($\beta$ is the reciprocal
temperature), $T_{\tau=0}=T_{\tau=\beta}$, and $\Delta\tau\rightarrow 0$ as
$N\rightarrow \infty~$.
Expanding $e^{-\Delta \tau H}$ and $T_{\tau +\Delta \tau}$, one obtains:
\[ \begin{array}{rl}
\langle{}0|T^+_{\tau_i}(1-\Delta \tau H)(T_{\tau_i}+\Delta \tau
   \partial_{\tau_i})
   T_{\tau_i}|0\rangle{} &
   \approx
   \langle{}0|T^+_{\tau_i} [1+\Delta \tau (\partial_{\tau} - H)]T_{\tau_i}
   |0\rangle{} \\[0.3cm]
   & \approx
   \langle{}0|T^+_{\tau_i}T_{\tau_i}|0\rangle{} \exp \left\{ \Delta \tau
   \frac{\langle{}0|T^+(\tau_i)(\partial_{\tau}-H)T(\tau_i)|0\rangle{}}
        {\langle{}0|T^+(\tau_i)T(\tau_i)|0\rangle{}} \right\}~,
\end{array} \]
and the partition function acquires the form
\[
Z=\int \prod_i (d\mu_i \, \langle{}0|T^+_{\tau_i}T_{\tau_i}|0\rangle{})
  \prod_i \exp \left\{ \Delta \tau

\frac{\langle{}0|T^+_{\tau_i}(\partial_{\tau}-H)T_{\tau_i}|0\rangle{}}
               {\langle{}0|T^+_{\tau_i}T_{\tau_i}|0\rangle{}} \right\}~,
\]
\[
N\rightarrow \infty~, ~~~\Delta \tau \rightarrow 0~.
\]
This equation is just the definition of a functional integral over
corresponding group parameters:
\begin{equation}
Z  = \int d\tilde{\mu} \exp \left[ \int^{\beta}_0 d\tau
  \frac{\langle{}0|T^+(\tau)(\partial_{\tau}-H)T(\tau)|0\rangle{}}
       {\langle{}0|T^+(\tau)T(\tau)|0\rangle{}}  \right]~.
\end{equation}

Wiegmann$^{6}$ has suggested a representation for the $t$-$J$ model
 in terms of the following parametrization
\[
P=\frac{1}{2} (1-\rho)(1+\vec{m}\vec{\sigma})_{\sigma\sigma'}
  X^{\sigma\sigma'} + \Psi_{\sigma}X^{\sigma 0}+
  X^{0\sigma}\Psi^+_{\sigma} + \rho X^{00}~,~~~
\rho=\sum_{\sigma} \Psi^+_{\sigma}\Psi_{\sigma},
\]
with the constraint $(1+\vec{m} \vec{\sigma})_{\sigma \sigma'} \Psi_{\sigma'}
=0$. Thus, only one Fermi variable exists.
The exchange interaction
is split with the help of the Hubbard-Stratonovich intermediate field
$\Delta_{ab} \equiv(\sigma_{ab},\pi_{ab})$, yielding the action
\begin{equation}
L=\sum_a u^+_a (i\partial_{\tau}+A_{\tau}+\mu)u_a +
  \sum_b v^+_b (i\partial_{\tau}+A_{\tau}-\mu)v_b
  -H_{ex} - H_t~,
\end{equation}
\begin{equation}
H_{ex}= \sum_{\langle{}ab\rangle{}} \{\sigma_{ab}u^+_av_b (Z^*_aZ_b) + \pi_{ab}
v^+_av_b
(Z_a\wedge Z_b) + h.c.\} +
\sum_{\langle{}ab\rangle{}} \frac{|\sigma_{ab}|^2 + |\pi_{ab}|^2}{2J_{ab}}~.
\end{equation}
The form (11), where $u_a=\Psi_a$, $v_b=\Psi_b$ ($a$, $b$ are  sublattices),
contains the same ``gauge'' terms originating from the spin wave
function as in the Batyev probe wave function. The second gauge field comes
from the Hubbard-Stratonovich field $(\sigma ,\pi )$.
Then, in$^{6}$ the general form of the exchange potential has been introduced:
\begin{equation}
W=\sum_C w(L_C) |\prod_C \Delta| \cos F(C)~,
\end{equation}
where
\[
\prod_C \Delta= \Delta_{ij}^* \Delta_{jk} \ldots \Delta_{ni}
\]
is
over a closed loop on the lattice. We find that $w(L_C)$ depends on the length
of $C$
and $F(C)={\rm Im}( \ln \prod_C \Delta)$ is the flux of the gauge field
through $C$.
 The flux state can arise due to fluctuating exchange bonds.
However, it should be emphasized that the general exchange potential $W$
may arise only via fermion-``boson'' loops which  are all to be taken
{\em at the same ``time''}.
It supposes the opportunity to neglect all dynamical contributions,
which is certainly questionable. An alternative would lie in the transformation
of the Hamiltonian to  collective variables describing these
loops. However, for a formulation of a functional representation these
variables must  form a complete set. If they are not orthogonal, one has to
work with
${\langle{}0|T^+(\tau)T(\tau)|0\rangle{}}$ in the measure and the exponential
in (9).\\
An investigation of the dynamical contributions, which always appear in
the usual
perturbation theory, requires a  recipe for the evaluation of these
corrections.
The latter can be done only after an accurate determination of
the measure of the
functional integral corresponding to the concrete parametrization.
We shall make an attempt to elucidate this question,
which is especially important for the slave-boson representation$^{8,9}$,
in the next section.

Another widely-used approach is that of projected fermions$^{10-13}$.
This approach can be used only for zero temperature. We emphasize that
it  {\em loses contributions of the kinematic interactions}.
In this approach, all order parameters are expressed in terms of the
same Fermi operators$^{13}$, namely
\[
|\chi| e^{i\Theta_{ij}}=\sum_{\sigma}
\langle{}c^+_{i\sigma}c_{j\sigma}\rangle{}
\]
for the ``flux''-phase, and
\[
\Delta_{ij}=\langle{}c_{i\uparrow}c_{j\downarrow} -
c_{i\downarrow}c_{j\uparrow}\rangle{}
\]
for the pairing amplitude of the $d$-wave RVB state.
For both amplitudes  several possibilities exist:\\
i)~~~Both ``operate'' in the spin subspace. The constraint is:
$\sum _{\sigma }c^+_{i\sigma}c_{i\sigma} = 1$. The Hamiltonian does not contain
a hopping term, and it respects the local gauge transformation
$c^+_{n\sigma} \rightarrow c^+_{n\sigma}e^{i\theta _{n}}$. Then, both order
parameters are locally
gauge non-invariant and equal to zero, if $<...>$ means full averaging as in
 (9). Actually, the Hamiltonian is invariant with respect to the local gauge
transformation\\
\begin{equation}
U{\cal H}U^{\dagger } = {\cal H},~~  U = U_{n}U_{m},~~ U_{n} = exp[-i\theta
_{n}\sum _{\sigma }c^{\dagger}_{n\sigma }c_{n\sigma }]. \nonumber
\end{equation}
With $U c_{n} U^{\dagger } = e^{-i\theta _{n}}c_{n}$, it follows that
\begin{equation}
\langle{}c^{\dagger }_{n\sigma }c_{m\sigma }\rangle{} = \frac{1}{Z} Sp\{
e^{-\beta {\cal H}} c^{\dagger }_{n}c_{m} \} =
\frac{1}{Z} Sp\{ e^{-\beta \cal H} U^{\dagger }Uc^{\dagger }_{n}U^{\dagger }
Uc_{m} U^{\dagger }U\} = e^{i(\theta _{n} - \theta _{m})}\langle{}
c^{\dagger }_{n\sigma }c_{m\sigma }\rangle{}, \nonumber
\end{equation}
or
\begin{equation}
(1 - e^{i(\theta _{n} - \theta _{m})})\langle{}c^{\dagger }_{n\sigma
}c_{m\sigma }\rangle{} = 0. \nonumber
\end{equation}
Since $(\theta _{n} - \theta _{m})$ is arbitrary,
 $\langle{}c^{\dagger }_{n\sigma }c_{m\sigma }\rangle{} = 0 $.\\
ii)~~Both ``operate'' in the charge channel. The constraint is
$~c^+_{i\sigma}c_{i\sigma}c^+_{i,-\sigma}c_{i,-\sigma} = 0$,
$c^+_{i\sigma}c_{i\sigma}$ can be equal to $1,0$. The hopping term in the
Hamiltonian and both order of the parameters are non-invariant
with respect to local gauge transformations. Both $\chi _{ij}$ and
$\Delta _{ij}$  can exist, but {\em must} disappear in the limit of zero
hopping amplitude. \\
iii)~Mixed case. This is possible only, if we use Fermi operators together
with non-exact constraints$^{10-17}$, or in case of systems with phase
separation.

Case i) corresponds to a non-local order parameter, which cannot be
described in the form of a simple product of single-site spin
wave functions.
Corresponding many-particle wave functions should provide the existence
of {\em single-time} loops of $\chi_{ij}$ or $\Delta_{ij}$, which
are not equal to zero.\\
The main idea in Refs.12-17 on the $t-J$-model is to introduce a new type of
magnetic state, which can be characterized by non-zero flux  of a fictious
magnetic field in each  super-cell of a 2D lattice. This field originates
from the exchange interaction and has to guide the motion of holes
(if there are) in such a way $^{12-15}$ that the exchange energy is not
much changed. This can be done by optimizing the flux for each particular
value of filling. This description is of the third type. However, it follows
from the analysis given here and in  Sec.2, that in systems with non-zero
concentration of holes, the description of the $t-J$ model in terms of
complex hopping amplitude (as in $^{12-17}$) is only possible in the
charge channel. To find these amplitudes, one must construct  self-consistent
equations for both the real and {\em the imaginary} parts. But errors
are easily introduced if the {\em local} constraints for the number of
available states are treated inexactly. This leads to mean-field theories
in terms of gauge non-invariant "order parameters"(see Sec.2). Therefore, we
find that  it is necessary, {\em first} to respect local constraints and
{\em then} to transform to momentum space$^{18}$, but not {\em vice versa},
as  has been done in $^{12-17}$. This  should also be done for any other
representation
 based on wave function "embracing" more than one unit cell. A probe
 many-electron wave functions may give non-zero
 $\langle{}c^{\dagger }_{n\sigma }c_{m\sigma }\rangle{} $ or
$\langle{}c^{\dagger }_{n\sigma }c_{m\sigma }e^{i\theta_{ij}}\rangle{}$.
However, as follows from Eqs.(13-15), an infinite number of other
many-electron wave functions should exist and be admixed with the initial one
providing the property
 $ \langle{}c^{\dagger }_{n\sigma }c_{m\sigma }\rangle{} = 0,
 $  if $[U,{\cal H}] = 0$. In the next section we reexamine
different functional representations of the partition function,
paying particular attention to the question of constraints. In Sec.3
we construct a mean field theory for the simplest flux-phase  state$^{16-17}$,
which, however, coexists with  magnetic order of Ne\'el-type.

\newpage
\subsubsection*{2. Parametrizations and constraints}
 It follows  from Eqs.(7) and (8), that the correct formulation of the
partition function using functional integrals requires  $^{6}$:\\
a)~ construction of  the projection operator $P=T|0\rangle{}\langle{}0|T^+$;\\
b)~evaluation of the measure $d\tilde{\mu}$.\\
The requirements for $T|0\rangle{}$ and for $d\mu$ are
\begin{equation}
\langle{}0|T^+T|0\rangle{}=1~,
\end{equation}
\begin{equation}
\int d\mu \, T|0\rangle{}\langle{}0| T^+ =\hat{1}~.
\end{equation}
Equation (17) is just the operator expansion of unity, which is needed
to construct (9).

This approach  reproduces the usual functional representation
for boson and fermion systems.
For bosons it is discussed in Ref. 7.
Let us now consider  different possible parametrizations of the
$0$-$\sigma$-model, i.e. the model with 3 states on each site:
$|0\rangle{}$, $|\uparrow\rangle{}$, and $|\downarrow\rangle{}$.
This model corresponds to the Hubbard model with $U=\infty$.
We start with the
representation used in Ref.6 and subsequent works $^{15}$. Let us put

\[
\begin{array}{rl}
      P & = T|0\rangle{}\langle{}0|T^+ \\
[0.2cm] & = [1+\frac{1}{2}\Psi^+\Psi+\Psi^+
                 Z_1X^{\uparrow 0}
                 +\Psi^+Z_2X^{\downarrow 0}]|0\rangle{}\langle{}0|[1+
                 \frac{1}{2}\Psi^+\Psi+Z_1^*\Psi X^{0\uparrow}+
                 Z_2^*\Psi X^{0\downarrow}]~,
\end{array}
\]
\begin{equation}
\langle{}0|T^+T|0\rangle{}=1+\Psi^+\Psi+\Psi^+\Psi Z^+Z=
1+\Psi^+\Psi(1-Z^+Z)=1~,
\end{equation}
\[
Z^+=(Z^*_1, Z^*_2)~,
\]
which guarantees that the product of spinors is $Z^+Z=1$ (this is the
so called $CP^1$-representation for the SU(2) group).
The parametrization of Hubbard operators yields:
\[
\langle{}0|T^+X^{00}T|0\rangle{}= 1+\Psi^+\Psi \equiv 1-\rho~,\]
\[
\langle{}0|T^+X^{\sigma 0}T|0\rangle{} = Z^*_{\sigma}\Psi~,\]
\begin{equation}
\langle{}0|T^+X^{0\sigma}T|0\rangle{}= \Psi^+Z_{\sigma}~,
\end{equation} \[
\langle{}0|T^+X^{\sigma \sigma'}T|0\rangle{}=
                           \Psi \Psi^+ Z^*_{\sigma} Z_{\sigma'} \equiv \rho
                           Z^*_{\sigma} Z_{\sigma'}~,
\]
and coincides with that  used in Refs.6 and 15.
In order to  define the measure, one has to choose a parametrization
for $Z_{\sigma}$.
Let us consider the following one:
\[
Z_1= \cos \Theta e^{i\varphi}~, ~~~Z_2= \sin \Theta e^{i\alpha}~.
\]
Then,
\begin{equation}
\int d\mu = \int d\Psi d\Psi^+ \frac{1}{2\pi^3} \int^{2\pi}_0
            d\varphi \int^{2\pi}_0 d\alpha \int^{\pi}_0 d\Theta
\end{equation}
and
\begin{equation}
\int d\mu \, T|0\rangle{}\langle{}0| T^+= X^{00} + X^{\uparrow \uparrow}
                                 + X^{\downarrow \downarrow} =\hat{1}~.
\end{equation}
Finally,
\begin{equation}
\langle{}0|T^+\partial_{\tau}T|0\rangle{}= \Psi^+\partial_{\tau} \Psi +
                           \rho Z^*_{\sigma}\partial_{\tau}Z_{\sigma}=
                           \Psi^+\partial_{\tau} \Psi +
                           \rho[\cos^2 \Theta \, i\partial_{\tau}\varphi+
                                \sin^2 \Theta \, i\partial_{\tau}\alpha]~.
\end{equation}
Here, the measure of integration is very simple for fermions, but the
integration over the nonlinear field $Z$ causes serious problems.
Up to now, these problems have not been solved even for the ``free'' field
and, hence, the correct statistics with the
``time''-dependent term (22)  cannot be reproduced even for noninteracting
particles.
 The best that can be done is to use a probe
wave function which allows us  to avoid time-dependent contributions.
But this means that
the mean-field theory is  constructed with
wrong statistics for the particles, and hence that
 the $T\neq 0$-region is unattainable (at least from
the viewpoint of correct statistics). In particular, neither Bose-condensation
nor  averages $<Z_{n}>$ for the field $Z_{n}$ are permitted.
Note that the parametrizations $Z_{\uparrow}=1$, $Z_{\downarrow}=\pm 1$
and $Z_{\uparrow}=Z_{\downarrow}^*=\exp (i\theta)$ suggested by Weller $^{19}$
cannot be used, since they do not describe the spin properties of the system
$^{20}$.
It is easy to show that $s^z=X^{\uparrow \uparrow}-X^{\downarrow \downarrow}=0$
always holds in this representation.

At this point we should mention --- in connection with the extension$^{13,14}$
of the
Azbel-Hofstadter problem of electrons in a plane in an
external magnetic field to the problem of strongly correlated
electrons --- that, in contrast to the Batyev representation (4), the
hopping term in the parametrization (19),
\begin{equation}
\sum_{\langle{}nm\rangle{}} t_{nm} \Psi^+_n \Psi_m Z_nZ^*_m~,
\end{equation}
 represented in the form
\[
t_{nm} \exp \left\{ i\int^m_n \vec{A} d\vec{l} \right\} \Psi^+_n\Psi_m~,
\]
cannot lead to a non-trivial phase (and flux) on any loop.
Suppose, for example, that the spin wave function of neighbouring
sites, $n$ and $m$, contains some
admixture of N\'eel type ordering, but is allowed to posses arbitrary
phases at each site:
\begin{equation}
Z_n= \left( \begin{array}{c} u e^{i\varphi_n}\\ v e^{-i\alpha_n} \end{array}
\right)~,~~~
Z_m= \left( \begin{array}{c} v e^{i\varphi_m}\\ u e^{-i\alpha_m} \end{array}
\right)~,
\end{equation}

\[
Z_1Z_2^* = 2uv\cos (\Phi^+_1-\Phi^+_2) e^{i(\Phi^-_1-\Phi^-_2)}~,
\]
\[
Z_2Z_3^* = 2uv\cos (\Phi^+_2-\Phi^+_3) e^{i(\Phi^-_2-\Phi^-_3)}~,
\]
where $\Phi^{\pm}=\alpha\pm\varphi$.
Then, any closed loop gives
\begin{equation}
\propto \prod_{ij} \cos (\Phi^+_i-\Phi^+_j)(2uv)^n~,
\end{equation}
{\em i.e.}\\
\[
\oint {\bf A}d{\bf l} \equiv 0.~
\]
 This shows that any spin wave function in the form of a simple product of
 single-site wave functions cannot provide a non-trivial phase factor for the
 hopping term. For more complicated wave functions the question is still open.
 The gauge field in Batyev's work$^{5}$ has appeared because of the expansion
and cannot provide a non-trivial flux through a plaquette.
Thus, these phases  describe only the type of magnetic ordering and the
dependence of the hopping term on it. For example, the expression (23)
shows explicitely that
for the ferromagnetic state $Z_nZ^*_m=1$ holds, and also that the bandwidth is
maximal.
For the N\'eel-type state, where $u=1$ and $v=0$ in (24), (23) gives zero,
corresponding to complete localization. Hence in this case the bandwidth is
determined by spin fluctuations.
Finally, minimization of the energy with respect to
$(u,v)$ in (24) gives also
the solution $u=v=1/\sqrt{2}$, yielding the maximum possible weight in
the loops (25), $2uv=1$.

Recently, Barnes$^{21}$ has suggested that the flux-phase state can be
provided only by the hopping term ({\em i.e.} at $J = 0$). His
investigation$^{21}$ was done by using the slave-boson approach, where
Hubbard operators are expressed in the form:

\begin{equation}
X^{00}=b^+b~,~~~X^{0\sigma}=b^+f_{\sigma}~,~~~
X^{\sigma 0}=f^+_{\sigma}b~,~~~X^{\sigma \sigma'}=f^+_{\sigma}f_{\sigma'}~.
\end{equation}
In order to reproduce (26), we have to parameterize $T$ in the form
\[
T|0\rangle{}=b|0\rangle{}+f_{\sigma}|\sigma\rangle{}~,
\]
\begin{equation}
\langle{}0|T^+=\langle{}0|b^++\langle{}\sigma|f^+_{\sigma}~.
\end{equation}
Then,
\[
\langle{}0|T^+T|0\rangle{}=b^+b+f^+_{\sigma}f_{\sigma}~,
\]
which implies the constraint
\begin{equation}
Q_{n}=b^{+}_{n}b_{n}+f^{+}_{n\sigma}f_{n\sigma}=1~.
\end{equation}
In the next step, we should provide the correct expansion of unity,
\begin{equation}
\int d\mu \,
T|0\rangle{}\langle{}0|T^+=
\int d\mu \, (b^+bX^{00}+ f_{\sigma}b^+X^{\sigma 0}
                           +X^{0 \sigma}bf^+_{\sigma}
                           +f_{\sigma}f^+_{\sigma'}X^{\sigma \sigma'})~.
\end{equation}
The measure $d\mu$ should contain $df_{\uparrow}df^+_{\uparrow}
df_{\downarrow}df^+_{\downarrow}$, and in the absence of any factor $F(f,f^+)$
in the first term ($\sim b^+b$) in (29), this four-operator term including
$X^{00}$
disappear according to the definition of integrals over Grassmann variables.
But this is wrong, since we must obtain $X^{00}+X^{\uparrow
\uparrow}+X^{\downarrow \downarrow} = 1$.
Thus, we are forced to add the factor $F=\exp\{-f^+_{\uparrow}f_{\uparrow}
-f^+_{\downarrow}f_{\downarrow}\}$.
The terms with $ X^{0\sigma}$, $X^{\sigma 0}$ give zero (odd number
of Grassmann variables).
Finally, we should provide convergency of the integral over the
``Bose''-field $b$:
\begin{equation}
d\mu= db db^+ df_{\uparrow} df^+_{\uparrow} df_{\downarrow} df^+_{\downarrow}
      (i/2\pi) \exp\left\{-\left[\sum f^+_{\sigma}f_{\sigma} + b^+b\right]
                                                                   \right\}~.
\end{equation}
If we were to use the constraint (28) before carrying out the integration,
and  put $Q=1$ in the exponential in (30), then we  would loose the
expansion of
unity in (29).
Otherwise, we will get both the denominators in the exponential of (9) and the
exponential in the measure (30). This means that one has to use some
projection: we must calculate every graph and then remove all wrong
contributions.
Note also, that the calculation with a weight factor in the measure
is extremely inconvenient since the weight factor is diagonal in
real space (``time''), but  $f^+_{\sigma}\partial_{\tau}f_{\sigma}$ is diagonal
in momentum (frequency) space.

Finally,there is the possibility of using to use some projection procedure.
Let us
postulate for the Green function, $\langle{}TX^{0\sigma}_n(\tau)
X^{\sigma 0}_m(\tau)\rangle{}$,  the following projection in the
slave-boson representation,
\begin{equation}
\begin{array}{rl}
G & = \langle{}TX^{0\sigma}_n(\tau) X^{\sigma 0}_m(\tau')\rangle{}
   = \langle{}Tb^+_n(\tau)f_{\sigma_n}(\tau)
   f^+_{\sigma_m}(\tau') b_m(\tau')\rangle{} \\[0.4cm]
  & = \frac{\lim_{L\rightarrow \infty} \prod_j \frac{1}{2L}
         \int^L_{-L} dp_j \exp\{-ip_j\beta\} \int D(b,b^+)D(f,f^+)
                          b^+_n(\tau)f_{\sigma n}(\tau)f^+_{\sigma m}
                          (\tau')b_m(\tau') \exp\{{\cal L}+{\cal L}_0\}}
        {\lim_{L\rightarrow \infty} \prod_j \frac{1}{2L}
         \int^L_{-L} dp_j \exp\{-ip_j\beta\} \int D(b,b^+)D(f,f^+)
                                          \exp\{{\cal L}+{\cal L}_0\}}~,
\end{array}
\end{equation}
where
\[
{\cal L}_0 = i \sum_n p_n Q_n~,~~~
Q_n=(b^+_nb_n+f^+_{n\sigma}f_{n\sigma})~.
\]
(This was used  in Refs.8, 9, but without projection).
All wrong contributions are killed by projecting on every site and,
say, for a system without interaction (31), gives the correct result
\[
G=T\sum_{n} \exp\{-i\omega_n(\tau-\tau')\}
  \frac{N_{\sigma}+N_0}{-i\omega_n+(\epsilon_{\sigma}-\epsilon_0)}~,
\]
\[
N_0=Z_0^{-1} \exp\{-\beta \epsilon_0\}~~~,N_{\sigma}= Z_0^{-1} \exp\{-\beta
\epsilon_{\sigma}\}~,
\]
\[
Z_0=\exp\{-\beta \epsilon_0\}+\exp\{-\beta \epsilon_{\uparrow}\}
                         +\exp\{-\beta \epsilon_{\downarrow}\}~.
\]
Here $\epsilon_0$ and $\epsilon_{\sigma}$ are the energies of the states
$|0\rangle{}$ and $|\sigma\rangle{}$, respectively.
Note that any local energy level is fluctuating
before the projection procedure is done: $\epsilon_{\alpha}=
\epsilon_{\alpha} + ip_j$, and the system cannot be described in
${\bf k}$-space. Since the interaction with the local gauge field does not
contain any small parameter, it {\em must} be taken into account first,
and then interactions responsible for the coherence of the state allow to
use the  $\vec{k}$-representation.\\

For completeness, we mention the Barnes-Zou representation $^{22,23}$ for the
case of a finite value of the Hubbard $U$,
\[
c_{n\sigma } = e^{\dagger}_{n}s_{n\sigma} + \sigma s^{\dagger
}_{-\sigma}d_{n},~~ X^{0\sigma}_{n} =  e^{\dagger}_{n}s_{n\sigma},~~
 X^{-\sigma,2}_{n} = \sigma s^{\dagger }_{-\sigma}d_{n},
\]
with the constraint
\[
Q_{n} = e^{\dagger }_{n}e_{n} + \sum _{\sigma } s^{\dagger}_{n,\sigma}
s_{n,\sigma} + d^{\dagger }_{n}d_{n} = 1.
\]
The expansion (26) for $T$  must be slightly changed for this parametrization,
thus
\[
T |0\rangle = [e + \sum_{\sigma }s_{\sigma }X^{\sigma ,0} +
d X^{2,0}]|0\rangle .
\]
All of the difficulties of using the slave-boson representation (28)-(30)
are also present here, since the measure of integration contains the
exponential
factor $exp\{-\sum_{n}Q_{n}\}$. Using projection (31) leads to the correct
expressions for the bare population numbers
\\ $N_{i} = exp\{-\beta(\epsilon _{i} - n_{i}\mu \}/\sum _{i}
exp\{-\beta(\epsilon _{i} - n_{i}\mu \}$, where $\epsilon _{i} =
 \epsilon _{0}, \epsilon _{\sigma}, \epsilon _{2}$
 ($\epsilon _{2} = \epsilon _{\downarrow} +  \epsilon _{\uparrow} + U$),
and  $n_{i}$ is the number of localized electrons in the state
$|i\rangle $ of an ion. Again, the Green functions like
$\langle e_{n}(\tau ) e^{\dagger }_{n}(\tau ')\rangle $ and the corresponding
averages are equal to zero due to the projection and only the terms
$\langle A_{n}(\tau ) A^{\dagger }_{n}(\tau ')\rangle $
with $A_{n} = X^{0\sigma }_{n}, X^{-\sigma ,2}$ are non-zero. It follows also
 that $s_{n\sigma }$ and $e_{n},d_{n}$ cannot be interpreted as neutral
 fermions and charged bosons respectively, as was done in Ref. 27,
 since electric neutrality of the crystal is given by the relation
\[
Z_{ion}^{eff} = 1\cdot \sum_{\sigma}N_{\sigma } + 2\cdot N_{2} =
1 + N_{2} - N_{0}.
\]
A charge in this system can be transfered only by $X^{0\sigma}$ and
$X^{\sigma 2}$ ($e^{\dagger }s_{\sigma }$ and $s^{\dagger }_{\sigma }d$).
The false conclusion about the  current being carried only by $e$-
and $d$-particles is based
on an  inadequate account of the constraints.

  We have  inspected thoroughly  the diagram corrections in the perturbation
  theory for (31) and its analogs for many-level systems. The result is
  that the
 projection technique is {\em exactly equivalent} to the
usual diagram technique for Hubbard operators $^{24}$. We conclude that the
projection technique is not more advantageous than the use of  $X$-operators.
We shall not present here this tedious analysis. The main point, however,
should be emphasized: after the projection, {\em there are no graphs that
arise} for pure "Bose" or "Fermi" Green functions connecting different sites.
All of these contributions are eliminated by the projection because they are
locally gauge non-invariant.
The practical prescription for readers wishing to
use slave-boson technique with the projection (31) consists of the following.
In the first step one performs a summation over Matsubara
frequencies in the expressions relating to every site. In the second step
projection
should be performed. This removes all terms containing $ip_{n}$ in the
denominator. Fermi or Bose functions arising after summation give
$exp\{-\beta \epsilon _{\alpha }\}$, which, together with the denominator
$Z_{0}$ of (31),  yield ion population numbers.\\

  In summary, we find that:\\
I. The main problem in using a gauge representation for the Hubbard operators
is related to  the uncertainty of defining the $\delta $-function $\delta
(Q_{n}-1)$ for the Grassmann variables. This leads to the necessity to
retain the projection factor
$<0|T^{\dagger }_{n}(\tau )T_{n}(\tau )|0>$ at every site $n$ and for
each "time" point
$\tau $ in the measure of integration $d\mu _{n}(\tau )$, while the action
of the system contains
$<0|T^{\dagger }_{n}(\tau )\partial _{\tau }T_{n}(\tau )|0>$ which is
diagonal in ``frequency"-space. Taking account of this projection unavoidably
leads to the diagram technique for Hubbard operators.\\
II. For the case of Hubbard $U = \infty $, the  suggestion in Ref.21 for
the existence of a flux-phase in the $t-J$-model
with $J=0$ (at least in the construction used in$^{21}$) cannot be accepted,
because the solution contains {locally non-invariant magnitudes}
$<b_{n+1}b^{\dagger }_{n}>$ ( $n$ are sites): after projection even loops of
these values are vanished and  only
$<f^{\dagger }_{n+1}b_{n+1}b^{\dagger }_{n}f_{n}> \neq 0$ are permitted.
The proof is analogous to that one given in (13)-(15), but with
$U_{n} = exp\{i\theta _{n}Q_{n}\}$,
$Q_{n} = b^{\dagger }_{n}b_{n} + \sum_{\sigma }f_{n\sigma}^{\dagger }
f_{n\sigma}$.\\
III. It is the phase of the Hubbard-Stratonovich field used for decoupling
of the exchange interaction which is responsible for the existence of
flux-phases. These phases are {\em independent} variables of spin wave
functions, hence {\em two order
parameters} can be introduced: the effective hopping amplitude, and the
magnetization.
It follows  that the order parameter
$A_{nm} = \langle c^{\dagger }_{n\sigma }c_{m\sigma }
exp\{i\theta _{nm}\}\rangle $ for the flux-phase ($c_{n}$ are fermions),
introduced in Ref.14 for minimizing the exchange energy, is equal to zero
for zero hopping, since in this case $[U_{n},{\cal H}_{J}] = 0$.
Hence, the mean field phase $\theta _{nm}$ cannot absorb {\em any, arbitrary}
and independent of  $\theta _{nm}$, difference of phases
$\theta _{n} -  \theta _{m}$ of the gauge transformation  $U_{n}U_{m}$ $^{25}$.

 In fact, a certain choice of $\theta _{nm}$ by definition of the  mean
 field violates {\em global} symmetry and, hence, corresponds to a finite
 number of variational parameters, while  an infinite number of variables
 $\theta _{n}$ describes {\em local} gauge symmetry, which cannot be
 violated by a phase transition.\\
 IV.  To correctly take  into account    statistics and kinematic
 interactions on a
 short-range scale any large-scale
action of the system should be derived from microscopics in terms of
Hubbard operators or by using many-particle Fadeev equations.

\newpage
\subsubsection*{3. Mean field equations}
 We now construct the mean field equations
for the $t$-$J$-model in terms of Hubbard operators:
\begin{equation}
H=H_0+H_t+H_J~,
\end{equation}
\[
H_0=\sum_i \Delta X_i^{\sigma \sigma}~,
\Delta=\epsilon_1^{\sigma}-\epsilon_0-\mu~,
\]
\[
H_t=\sum (t_{ij}X_i^{0\sigma}X_j^{\sigma 0} + H.c.),~
\]
\[
H_J=2\sum J_{ij}X_i^{\alpha \beta}X_j^{\gamma \delta}
         \vec{\sigma}_{\alpha \beta} \vec{\sigma}_{\gamma \delta},
\]
where $\epsilon_1$ and $\epsilon_0$ are the energies of single-
and double-occupied states of the ion (or, more correctly, states with
zero spin of an in-plane
elementary cell of a high-$T_{c}$-superconductor in the hole representation),
$t$ and $J$ denote the hopping and the exchange
integral, respectively, the $\mu$ is the chemical potential.
 Any site (lattice cell) has 3 states
only: $|0\rangle{}$, $|\uparrow\rangle{}$, or $|\downarrow\rangle{}$.

The present alternative formulation of a mean field theory manages to avoid
both the Hubbard-Stratonovich decoupling and any approximations in dealing
with
constraints like
\[
f^+_{i\sigma} f_{i\sigma} + b^+_i b_i = 1~~~{\rm or}~~~ z^+_{i\sigma}
z_{i\sigma} =1.
\]
Here, we consider the possible existence of a flux phase in the
``charge channel'' at the mean field level. Simultaneously we
examine the possibility of N\'eel-type ordering.

The procedure  is as follows.  First, we introduce the
Hamiltonians of effective hopping  $H_{\chi}$, and of  N\'eel field
$h_{\alpha }$ in sublattices $\alpha = A,B$,  $H_{h}$,
\begin{equation}
(H_0 + H_h) + H_{\chi}=  \sum_{i\alpha} (\Delta - h_{\alpha} \eta(\sigma))
               X^{\sigma \sigma}_{i\alpha} +
               \sum_{\langle ij\rangle,\langle \alpha \beta \rangle}
              [ \chi^{\alpha\beta}_{ij} X^{0\sigma}_{i\alpha}
               X^{\sigma 0}_{j\beta} + H.c.]
             ~,
\end{equation}
and solve the equations for the Matsubara Green function $\langle T
X^{0\sigma}_{i\alpha}(\tau) X^{\sigma 0}_{j\beta}(\tau') \rangle$
in the simplest approximation for the elementary cell of the lattice
shown in Figure 1. At this stage, the aim is to obtain the the single-particle
Green function with the well-known spectrum of the flux-phase$^{16,17}$.\\
The next step is the derivation of a self-consistent equation for $\chi$.
We obtain it from the requirement of a vanishing first correction
to the self-energy $\Sigma^{(1)}$ in the Green function from the
interaction
\begin{equation}
H_{int}=H_t+H_J-H_{\chi}-H_{h}~.
\end{equation}
Two further equations are needed: one for the magnetic
 order parameter and one for the dependence of the chemical
potential on the number of holes.

  The diagram technique for Hubbard operators is described e.g. in $^{24,26}$.
We use the following notations.
The bare Green function is
\begin{equation}
G^{(0)}_{\alpha} =
\langle TX^{0\sigma}_{i\alpha}(\tau) X^{\sigma 0}_{i\alpha}(\tau')
\rangle^{(0)} = T\sum_n e^{-i\omega_n (\tau-\tau')}P^{\sigma}_{i\alpha}
                /(-i\omega_n+\Delta^{(\alpha)}_{\sigma 0})~,
\end{equation}
where
\[
P^{\sigma}_{i\alpha} = N^0_{i\alpha} + N^{\sigma}_{i\alpha}~,
\]
\[
N^0_{i\alpha} \equiv \langle X^{00}_{i\alpha}\rangle~,~~~
N^{\sigma}_{i\alpha}\equiv \langle X^{\sigma \sigma}_{i\alpha} \rangle~,
\]
\[
\Delta^{(\alpha)}_{\sigma 0} = \epsilon^{\sigma}_{i\alpha} - \epsilon_0
                               - \mu = \Delta - h_{\alpha}\eta (\sigma)~,
\]
\[
\eta(\sigma =\uparrow)=1~,~~~\eta(\sigma =\downarrow)=-1~,~~~h_A=h~,
{}~~~h_B=-h~,~~~\omega_n=(2n+1)\pi T~,
\]
and $\alpha = A,B$ denotes the sublattice.
The diagrammatic notations$^{26}$ are
\begin{equation}
\begin{array}{ll}
\stackrel{-~-~-~-~-}{i\alpha~~~~~~j\beta} &
{}~~\Longleftrightarrow ~~\chi^{\alpha
                                         \beta}_{ij}~; \\[0.3cm]
\longrightarrow \!\!\! - &
                            ~~\Longleftrightarrow ~~T\sum_n e^{-i\omega_n(\tau
                            -\tau')}\frac{1}{-i\omega_n+\Delta^{(\alpha)}_{
                            \sigma 0}} \equiv D^{0\sigma}_{i\alpha}
                            (\tau -\tau')~; \\[0.3cm]
\longrightarrow\!\!\! -\bullet  &
                            ~~\Longleftrightarrow ~~G^{0\sigma}_{i\alpha}
                            (\tau -\tau') \equiv \langle TX^{0\sigma}_{
                            i\alpha}(\tau)X^{\sigma 0}_{i\alpha}
                            (\tau')\rangle^{(0)}~.
\end{array}
\end{equation}
The dot at the end of a Green function line denotes $P^{\sigma}_{i\alpha}$.
Using (35) and calculating the limit $\tau' \rightarrow \tau + \varepsilon$
($\varepsilon >0$, $\varepsilon \rightarrow 0$), one obtains
\begin{eqnarray}
\langle X^{\sigma \sigma}\rangle & \mbox{\hspace{-20mm}}= N_{\sigma} =
   - \lim_{\varepsilon \rightarrow 0} T\sum_n e^{i\omega_n \varepsilon}
   \frac{N_0+N_{\sigma}}{-i\omega_n +\Delta_{\sigma 0}}\nonumber \\
   & =
   -\frac{1}{2\pi i} \lim_{\varepsilon \rightarrow 0}
   ~{\displaystyle \int} ~dz~~
   \frac{e^{z\varepsilon}}{e^{\beta z}+1} \cdot
   \frac{N_0 +N_{\sigma}}{-z+\Delta_{\sigma 0}} =
   (N_0 +N_{\sigma})\frac{1}{e^{\beta \Delta_{\sigma 0}}+1}.
\end{eqnarray}

Equation (37) and the identity $N_0 + N_{\uparrow} + N_{\downarrow} =1$
give the correct expressions

\begin{equation}
N^{(0)}_0 = \frac{1}{1+e^{-\beta \Delta_{\uparrow 0}}+e^{-\beta \Delta_{
                     \downarrow 0}}}~;~
N^{(0)}_{\sigma} = \frac{e^{-\beta \Delta_{\sigma 0}}}{1+
                           e^{-\beta \Delta_{\uparrow 0}} + e^{
                              -\beta \Delta_{\downarrow 0}}}~.
\end{equation}
Hence, the correct statistics is obtained as is easily checked by a
direct evaluation of
$N_0 \equiv {\rm Sp}\{ X^{00} e^{-\beta H}\}/{\rm Sp}\{e^{-\beta H}\}$.
 The slave-boson, slave-fermion, and other
gauge theories (including $1/N$-expansion) fail to provide
the correct result (38) for $H=H_0$.\\
The consequent decoupling of Fermi-type Hubbard operators leads to a
$T_{\tau}$-average of a product of diagonal Bose-type operators.
The latter one is to be calculated with the help of the usual
cumulant technique.
The choice of $\chi$ and of the N\'eel magnetization
in the lattice are shown in Figure 1.

The mean field Hamiltonian (33) does not allow an exact solution
 for the Green functions,
because of the nonlinear nature of the Hubbard operators themselves.
Therefore we start with a ``chain'' approximation for the Green
functions, gathering the maximum number of pole contributions
(or, what is the same, accounting for the first correction to the
self-energy). The result is similar to the renormalized mean field
introduced in Ref.12,
however, with another renormalization factor.
In analytical form, the result is written as
\begin{equation}
G^{\gamma \delta}_{nm\sigma} = G^{(0)\gamma}_{n\sigma} \delta_{nm}
                               \delta^{\gamma \delta} +
                               G^{(0)\gamma}_{n\sigma} \chi^{\gamma \nu}_{
                               nl} G^{\nu \delta}_{lm\sigma}~,
\end{equation}
where summation over repeating indices is implied.

The Fourier transformation
$(\vec{R}_n-\vec{R}_m$, $\tau' -\tau) \longrightarrow (\vec{k}$,
$i\omega)$ leads to
\[
G^{\gamma \delta}_{\vec{k}\sigma} =
                  G^{(0)\gamma}_{\sigma} \delta^{\gamma \delta}+
                  G^{(0)\gamma}_{\sigma} \chi^{\gamma \nu}_{\vec{k}}
                  G^{\nu \delta}_{\vec{k}\sigma}
\]
or, in matrix form, to
\begin{equation}
\hat{G}^{-1}_{\vec{k}\sigma} = (\hat{G}^{(0)}_{\vec{k}\sigma})^{-1} -
\hat{\chi}_{\vec{k}\sigma}~,
\end{equation}
where the right-hand side matrix is
\begin{equation}
\left( \begin{array}{cccc}
       (G^{(0)}_A)^{-1} & E^-_{13x}      & 0              & (E^+_{42y})^*
       \\[0.2cm]
                      & (G^{(0)}_B)^{-1} & E^-_{24y}      & 0
       \\[0.2cm]
             ~~~c.c.       &           & (G^{(0)}_A)^{-1} & E^-_{31x}
       \\[0.2cm]
                         &               &                & (G^{(0)}_B)^{-1}
\end{array}
\right) = \hat{G}^{-1}_{\vec{k}\sigma}~.
\end{equation}
Here,
\[
E^-_{13x}=-(\chi_1+\chi_3 e^{-ik_x})~,
{}~~~E^{+}_{42y}=-(\chi_4+\chi_2 e^{+ik_y})~,~~
\]
{\em etc.}
The spectrum of the system,  given by det$\| G^{-1}_{\vec{k}\sigma}\|=0$, is
\begin{equation}
E^{\sigma}_{1,2,3,4}=(+i\omega)_{1,2,3,4}=\Delta\pm\sqrt{h^2+\epsilon^2_{
                     \pm}(k_x,k_y)}~,
\end{equation}
where
\begin{equation}
\epsilon^2_{\pm}(k_x,k_y) \equiv |\chi_1 e^{ik_x/2}+ \chi_3 e^{-ik_x/2} \pm
                           (\chi_2 e^{-ik_y/2} +\chi_4 e^{ik_y/2})|^2
                           P^{\sigma}_A P^{\sigma}_B~.
\end{equation}
We do not intend to search for a charge density wave here, so we put $N_{0A} =
N_{0B} = N_{0}$.
Since we consider a N\'eel-type of structure, $E^{\uparrow}_i(\vec{k})=
E^{\downarrow}_i(\vec{k})$.
Let us denote
\[
N^{\sigma}_A=\frac{1}{2}N_1+\eta (\sigma) m~,
\]
\[
N^{\sigma}_B=\frac{1}{2}N_1-\eta(\sigma) m~,
\]
\[
N_0+N_1=1~.
\]
Then,
\begin{equation}
P^{\sigma}_A P^{\sigma}_B = \left( \frac{1+N_0}{2} \right)^2 -m^2 \equiv
                            \left( \frac{1+\delta}{2} \right)^2 -m^2~.
\end{equation}
Here $\delta=N_0$ is the hole concentration. The total number of holes
in the plane is
\[
2 \cdot N_0+ 1 \cdot N_1 =\rho =N_{holes}/N_{sites} = N_0+1~.
\]
Hence, $\delta=\rho -1$ is the deviation from half-filling.
The equation
\begin{equation}
\delta=N_0
\end{equation}
is the (self-consistent) condition for the determination of the chemical
potential $\mu$.
In a system without interaction, the equation
\[
\delta=N_0=(1+2e^{-\beta \Delta})^{-1}
\]
yields $\mu =\omega =\epsilon_1 -\epsilon_0$ or $\Delta =0$ for the whole
interval $\delta \in (0,1)$ at $T=0$.
In a system with interaction, however, $\mu=\mu(\delta) \neq \omega =
\epsilon_1 -\epsilon_0$.
Putting $\Delta=h=0$ in (43), we find that the spectrum (42) almost
coincides with that found by Affleck and Marston$^{17}$.
The only difference is the factor $P_A P_B$ (equal to 1/4 at half-filling)
involved in equation (42).
In comparison with Ref.17, the corresponding factor amounts to 1/2.

Now consider the situation $\chi_i=|\chi|e^{i\Theta}$.
In this case, equation (43) becomes
\begin{equation}
E_{1,2,3,4}(\vec{k})=\Delta \pm \nu_{\pm}(\vec{k}),
\end{equation}
where
\[
\nu_{\pm} =
 \sqrt{h^2+|\chi|^2(|\gamma_x|^2+|\gamma_y|^2 \pm 2|\gamma_x| |\gamma_y|
                                \cos 2\Theta) P_AP_B}~,
\]
\[
\gamma_x = \gamma (k_x) = 1+e^{-ik_x}~,~~~
\gamma_y = \gamma (k_y) = 1+e^{-ik_y}~.
\]
The choice $\Theta=\pi/4$ leads to the ``flux-phase'' spectrum $\propto
\sqrt{|\gamma_x|^2+|\gamma_y|^2}$ discussed in Refs. 17,12.
There exist two possibilities: one can  minimize the free energy of
the system with respect to
$\Theta$, or one can search for different solutions, corresponding to
$\Theta=\pi/4,\pi/2,\pi$ --- i.e. to some choosen values.
The question arises, however, whether these values are consistent
with the equation for $\chi$.
Let us construct this equation, applying the Gor'kov--Nambu formalism
$^{27, 28}$ to our problem.
To do this  we have to calculate the first correction
to the Green function $\langle TX^{0\sigma}_{i\alpha}(\tau)
X^{\sigma 0}_{j\beta}(\tau') \rangle$ from $H_{int}$.
Explicitely,
\begin{eqnarray}
H_{int}& = &\sum_{ij\alpha \beta} [(t^{\alpha \beta}_{ij} X^{0\sigma}_{i\alpha}
        X^{\sigma 0}_{j\beta} + H.c.) + \nonumber  \\
       &   &(J^{\alpha \beta}_{ij}\vec{\sigma}_{s_1s_2}
        \vec{\sigma}_{s_3s_4}X^{s_1s_2}_{i\alpha}X^{s_3s_4}_{j\beta} + H.c.)
-\nonumber  \\
       &   &(\chi^{\alpha \beta}_{ij}X^{0\sigma}_{i\alpha}X^{\sigma
0}_{j\beta}+ H.c.) +
 h_{i\alpha}\sigma^z_{s_1s_2}X^{s_1s_2}_{i\alpha}\delta_{ij}
        \delta^{\alpha \beta}]~.
\end{eqnarray}
The correction we are interested in must contain exchange and hopping
in the denominator of the Green function in first order.

Thus, the equation for $\chi$ is
%\begin{figure}
\begin{picture}(100,10)(0,-100)
\put(56,-153){\oval(53,40)[t]}
\end{picture}
\begin{equation}
\begin{array}{ccccc}
                            &   &           \vec{k}          &   & \\[0.6cm]
\sim\!\sim\! \sim\! \sim\! \sim    & + & .................  & = & ---\, - \\
\alpha~~~~\vec{p}~~~~\beta  &   & \alpha~~~\vec{p}-\vec{k}~~~\beta &   &
{}~~~\alpha~~~~\vec{p}~~~~\beta~~,
\end{array}
\end{equation}
%\end{figure}
 or
\begin{equation}
t^{\alpha \beta}(\vec{p})+\Sigma^{(1)~\alpha \beta}(\vec{p}) = \chi^{\alpha
\beta}(\vec{p})~.
\end{equation}

Here $\sim\!\sim\! \sim\! \sim\! \sim $ denotes $t$ and
  $ .............$ denotes $J$.

The self-energy $\Sigma^{(1)}(\vec{p})$ is
\begin{equation}
\Sigma^{(1)\alpha \beta}_{\sigma}(\vec{p})=\frac{T}{N}
\sum_{\vec{i},\vec{h},\omega} J^{\alpha \beta}_{\vec{i},\vec{i}+\vec{h}}
G^{\alpha \beta}_{\sigma,\vec{i},\vec{i}+\vec{h}} e^{i\vec{h}\vec{p}} =
\frac{T}{N}\sum_{\omega_n \vec{k}}
G^{\alpha \beta}_{\sigma}(\vec{k},i\omega_n) J^{\alpha
\beta}(\vec{p}-\vec{k})~.
\end{equation}
For the case $\chi_i=|\chi|e^{i\Theta}$, the Green function in (48),
(50) is given by
\begin{equation}
G^{\alpha \beta}_{\sigma} (\vec{k},i\omega_n) =
A^{\alpha \beta}_{\vec{k}\sigma}/{\rm det}
\|G_{\sigma}^{-1}(\vec{k},i\omega_n))\|~,
\end{equation}
\vspace{0.2cm}
\[
G_{\sigma}^{-1}(\vec{k},i\omega_n) = \left(
\begin{array}{cccc}
(G^{(0)}_A)^{-1} & -\chi \gamma_x & 0               & -\chi^* \gamma_{y}
\\[0.2cm]
                 & (G^{(0)}_B)^{-1} & -\chi \gamma_y & 0 \\[0.2cm]
{}~~~c.c.          &               & (G^{(0)}_A)^{-1} & -\chi \gamma_x^*
\\[0.2cm]
                 &                &                 & (G^{(0)}_B)^{-1}
\end{array}
\right) ~,
\]
\vspace{0.1cm}
\begin{equation}
(P_A P_B)^2
{\rm det}\| (G_{\sigma}(\vec{k},i\omega_n))^{-1}\|=
\prod^4_{i=1}(-i\omega_n + E^{(i)}_{\vec{k}\sigma})
= [(-i\omega_n)^2 -\nu_+^2(\vec{k})][(-i\omega_n)^2 - \nu_-^2(\vec{k})]~.
\end{equation}
(Note that in the left-hand-side the factor $(P_A P_B)^2$ is present!).
The energy is measured from $\Delta$, and the matrix
$A^{\alpha \beta}_{\vec{k}\sigma}$ is
\begin{equation}
\hat{A}_{\vec{k}\sigma}=  \left(
\begin{array}{cccc}
(G^{(0)}_B)^{-1}d & \gamma_x\chi a & g_B \gamma_x \gamma_y
                                       & \gamma_y \chi c \\[0.2cm]
 & (G^{(0)}_A)^{-1}d & \gamma_y \chi c^* & g_A \gamma_x \gamma_y^* \\[0.2cm]
{}~~~~c.c. &  & (G^{(0)}_B)^{-1} d & \gamma^*_x \chi a \\[0.2cm]
 & & & (G_A^{(0)})^{-1} d
\end{array} \right)~,
\end{equation}
where we have used
\[
g_{\alpha} = 2|\chi|^2 \cos (2\Theta) (G^{(0)}_{\alpha})^{-1},
\]
\[
d(\vec{k},\omega)=
[(G^{(0)}_A)^{-1}(G^{(0)}_B)^{-1}-|\chi|^2(|\gamma_x|^2+|\gamma_y|^2)]~,
\]
\[
a(\vec{k},\omega)=[(G^{(0)}_A)^{-1}(G^{(0)}_B)^{-1}-
                   |\chi|^2(|\gamma_x|^2-e^{-4i\Theta}|\gamma_y|^2)]~,
\]
\[
c(\vec{k},\omega)=[(G^{(0)}_A)^{-1}(G^{(0)}_B)^{-1}+
                           |\chi|^2(e^{4i\Theta}|\gamma_x|^2-|\gamma_y|^2)]~.
\]
Consider the equation for the ``flux-phase order parameter'' $\chi$,
\begin{equation}
\chi\gamma (p_x) =
                  t\gamma (p_x)  + 2 \chi J \frac{T}{N}\sum_{\omega \vec{k}}
                  \gamma (p_x-k_x) \gamma_x \frac{
                  (i\omega)^2-h^2-|\chi|^2P_AP_B(|\gamma_x|^2-
                     e^{-4i\Theta}|\gamma_y|^2)}{
                  [(i\omega)^2-\nu^2_+][(i\omega)^2-\nu^2_-]}~,
\end{equation}
which is obtained from the matrix element $G^{12}_{\vec{k}\sigma}(i\omega)$.
Calculating the sum over Matsubara frequencies,
and separating the real and imaginary parts in equation (54), we find
the  pair of equations:
\begin{equation}
\begin{array}{l}
|\chi|\cos \Theta =
t - \frac{1}{2} |\chi| \cos \Theta \: \Phi_1~, \\[0.2cm]
|\chi|\sin \Theta = - \frac{1}{2}|\chi| \sin \Theta \: \Phi_2~,
\end{array}
\end{equation}
with
\[
\Phi_1 = \frac{1}{4N} \sum_{\vec{k}} (b_+(\vec{k}) l_+(\vec{k}) +
                                       b_-(\vec{k}) l_-(\vec{k}))~,
\]
\[
\Phi_2 = \frac{1}{4N} \sum_{\vec{k}} (b_-(\vec{k}) l_+(\vec{k}) +
                                       b_+(\vec{k}) l_-(\vec{k}))~,
\]
\[
b_{\pm}(\vec{k})=[|\gamma_x|\pm |\gamma_y|]^2~,~~~
l_{\pm}(\vec{k})=\frac{J}{\nu_{\pm}} [f(-\nu_{\pm})-f(\nu_{\pm})]~,~~~
f(x) = (1+e^{\beta x})^{-1}~.
\]
For $\Theta=0$, the second  equation in (55) disappears.
The equations for $\mu$ and $m$ are obtained from
\[
N^{(\alpha)}_{\sigma}=-\lim_{\tau'\rightarrow\tau+\varepsilon}
                      \left\langle TX^{0\sigma}_{i\alpha}(\tau)
                      X^{\sigma 0}_{i\alpha}(\tau') \right\rangle=
                      -\lim_{\varepsilon\rightarrow +0} T\sum_{\omega_n}
                      e^{-i\omega_n\varepsilon}G_i^{\alpha\alpha}(i\omega_n)~.
\]
Performing the  frequency summation, we obtain
\begin{equation}
N^{\sigma}_A=  (N^{\sigma}_A+N^0_A)
(n+ \eta (\sigma) \frac{h}{4J} L)~,
\end{equation}
\begin{equation}
n= \frac{1}{4N} \sum_{\vec{k}} [f(-\nu_+) + f(\nu_+) + f(-\nu_-) + f(\nu_-)]~,
\end{equation}
\begin{equation}
L= \frac{1}{N} \sum_{\vec{k}} [l_+(\vec{k}) + l_-(\vec{k})]~,
\end{equation}
where the sign of the staggered magnetization  is absorbed in $h$:
\[
\Delta^A_{\sigma 0}=\epsilon^{\sigma}_{A}-\epsilon^0_A-\mu=
                    \epsilon_1-\epsilon_0-\eta (\sigma)h-\mu=
                    \Delta -\eta (\sigma)h~.
\]
The N\'eel field, as usual, is $h=Jm$, where $m$ is now determined by
 (56).

We can write the complete system for $\mu$, $|\chi|$,
$\Theta$ and $m$ as
\begin{equation}
\left\{ \begin{array}{l}
0 = t- |\chi| \cos \Theta \; (1+\frac{1}{2}\Phi_1)  \\[0.2cm]
0 =    |\chi| \sin \Theta \; (1+\frac{1}{2}\Phi_2)  \\[0.2cm]
0 = m(1 - n - \frac{1+\delta}{8} L)      \\[0.2cm]
0 = \frac{1-\delta}{1+\delta} - n - \frac{m^2L}{2(1+\delta )} ~,
\end{array} \right.
\end{equation}
where $\Phi_1$, $\Phi_2$, $n$, and $L$
are given by the equations (55), (57), and (58). Switching off interaction
one finds from (57)
\[
N_0=\frac{1-n_0}{1+n_0}=\frac{1-\frac{1}{e^{\beta \Delta}+1}}{
                              1+\frac{1}{e^{\beta \Delta}+1}}=
                              \frac{1}{1+2e^{-\beta \Delta}}~,
\]
as it should be.\\
First, let us inspect the system (59) for possible flux-phase
solutions for the pure exchange interaction (the analogue of Refs.16, 17).
According to the  conclusions of Sec.2, such solutions should not exist.
It is easy to see that the functions $l_{\pm}({\bf k})\geq 0$. Therefore,
the functions $\Phi_1 \geq 0$ and  $\Phi_2 \geq 0$ and the second equation
(59) has only the
trivial solution $\Theta = 0$ for antiferromagnetic sign of the
exchange interaction. However, for ferromagnetic sign non-trivial
solutions also do not appear.
Let us suppose that a solution for  $t=0,~~\delta = 0,
 ~~\theta = \pi /4, m = 0$,  and $|\chi | \neq 0$ exists. Then the
 first two equations of  (59) give
\[
2 = -\Phi _{1} = -\Phi _{2}.
\]
The function $ \Phi _{1}$ is must be equal to $ \Phi _{2}$ for this set
of parameters, since $\nu _{+} = \nu _{-} = \nu $ and $l_{+} = l_{-} = l.$
However, when $\delta = 0$, which corresponds to a half-filled ordinary
band or fully filled lower strongly correlated band, both Fermi functions,
$f(-\nu_{+})$ and  $f(\nu_{+})$,
are equal to unity in every point of the Brillouin zone. One finds from
the definition of
 $l_{\pm}$ that $l({\bf k}) = 0$ holds and, therefore,
 $\Phi _{1} = \Phi _{2} = 0$. So, there are no flux phase solutions.
 It is also clear, that no solutions should exist
for $\delta = 0$ for any value of $\Theta $ either, since the term
$\chi _{ij}X_{i}^{\sigma 0} X_{j}^{0\sigma } $ deviates from zero only if
there are sites for particles  to hop to. It is  seen from Eq.(55) that
$ \Phi _{1} = \Phi _{2} = 0$ holds in this case. Therefore,
$\Theta  = 0,$ $|\chi |= t$. It follows  that this result is exact and
does not depend on the approximation used.  The conclusion is a
consequence of the constraint
$c^{\dagger }_{n}c_{n} = 1$, which is exactly taken into account by
using   Hubbard operators.

Let us take now $\Theta = 0$ and consider N\'eel antiferromagnetism (AF).
For
$t = 0$ one has $|\chi | = 0$, and the usual picture for mean field
N\'eel AF of localized electrons emerges. For $\delta \neq 0$, the Eqs.(59)
cannot be solved analytically.  N\'eel magnetization (see Fig.2) exists
in a finite interval of hole concentration and dissappears at
$\delta = \delta _{c}$. At $t = 0$, but  $\delta \neq 0$ (statistically
homogeneous distribution of empty sites in a Heisenberg AF) we have
the trivial mean field theory, and $m$ dissappears only at $\delta = 1$.
This is not correct, of course, since in this case the percolation problem
must be considered. However, this demonstrates  the nature of the mean
field used:
at a finite number of holes we spread the wave functions of correlated
electrons over the whole crystal and obtain Bloch functions.
In this approach exchange interaction between neighbouring sites
exist even at an infinitely small concentration of electrons, which is why
we obtain this strange result.
 Thus the results can reasonably be used in the region where
 $(1 - \delta )$ is not very small. When hopping is not equal to zero
 ($t \neq 0$),
 it opposes AF, and the larger the transfer term $t$, the less is the critical
 concentration $\delta _{c}$. As it is seen from the dependence of N\'eel
 magnetization on the hole concentration, shown in Fig.2,
 the character of solution dissappearance
 is rather non-trivial. At sufficiently large ratio $t/J$ there a second
 solution for finite
 magnetization arises, corresponding to larger energy of the
 system. The point where these solutions meet is a second critical
 value of $\delta$.
 Here, the magnetization jumps
 from a finite value to zero. This result
 cannot be extracted from the analysis
 of the paramagnetic susceptibility$^{26}$.
 The densities of states, corresponding to these branches of solutions, are
 shown in Fig.3: in the case, when the system has larger magnetization, only
 states below the antiferromagnetic gap are occupied, while in the case of
 smaller magnetization the states inside of the van Hove pecularity above the
 gap  are partly
 filled. The latter increases the energy of the system.
 In the limit $t/J \rightarrow \infty$, both critical concentrations
 tend to the same asymptotical value, $\delta_c(t\rightarrow \infty)=1/3$.
 The origin of this number can be understood from the last equation (59)
 for chemical potential. Let us take $t = J = 0$. Then we find from the
 equation
 \begin{equation}
 \frac{1 - \delta }{1 + \delta} = n
 \end{equation}
 that at $\delta = 1/3$ $n = 1/2$. At $t \neq 0$, $J \neq 0$ (see Eq.(57))
 only two lower bands are filled and the system has a gain in energy due to
 AF gap. A half of  available in ${\bf k}$-space states are filled in our
 tree-level system $(|0>, |\uparrow >, |\downarrow >)$ at $\delta = 1/3$.
 This value coincides with that one found in Ref.26. The dependence of
 $\delta _{c} (t)$ at zero temperature is shown in Fig.4.
The renormalization of hopping by exchange interaction does not
contain interesting effects: exchange supresses hopping in the
case of antiferromagnetic ordering, what we have seen already in the Sec.2
in terms of the
functional representation.

 \subsubsection*{Concluding remarks}
   In this work we have expressed doubts on the reliability of certain
   results for systems of strongly correlated electrons,
   which were obtained in the framework of
   various gauge theories.
  We have focussed particularly on the flux phases in the $t - J$-model.
  Our critisism is based on the difficulty of reconciling  the
existence of {\em exact} local gauge invariance of the Heisenberg or the
$t - J$ model  with averages (correlators and Green functions),
which do not take into account exactly the phase space reduction  at
{\em every} site. However, it should be emphasized, that this argument
applies only to certain types of flux phase description, but not to
the possible existence of the flux phase state itself. For example, it is easy
to check that
for a cluster of 4 sites (Fig.1) with one hole and  three spins, the
exact spin wave function
$|\Phi >$ of the chiral operator$^{29}$
$\chi _{123} = {\bf s_{1}\cdot (s_{2}\times s_{3})}$ is simulteneously an
eigenfunction of the operator
\begin{equation}
 \hat{T} = \hat{t}_{12} \hat{t}_{23} \hat{t}_{34} \hat{t}_{41}
\end{equation}
of charge transfer over the loop with hopping on the link $(nm)$
\[
 \hat{t}_{nm} = \sum _{\sigma } t_{nm} X_{n}^{\sigma 0}X_{m}^{0 \sigma }
+ H.c. ~.
\]

 If the spin part of the Hamiltonian can provide the appropriate spin
 wave function
(see the detailed discussion in Ref.29), one can hope that the
contribution of loop operators from the kinetic energy in the
ground-state wave function will be large enough to be seen. However,
this kind of flux phase  can appear {\em only at nonzero concentration
of holes}. The particular mean field description that we have used here
does not contain
this possibility. In order to do better, the mean field should be
constructed in terms of cluster variables or  eigenfunctions of
loop operators (60).

The  hope to construct such a system  experimentally seems to be
rather small.
 The physical nature of the flux states consists of the commensurability
 of two lengths: lattice spacing and ``cyclotronic radius" of the
 fictious magnetic field, coming from the spin current. This  obviously
 can arise only if the ratio of the parameters of exchange and hopping
is  of the order of unity. So, the following requirements must be
fulfiled:\\
(i)  the system must be two dimensional;\\
(ii) an exchange interaction must be not only large, but of
{\em antiferromagnetic} sign;\\
(iii) the hopping must couple the same neighbouring magnetic elementary
cells of the effective  lattice;\\
(iv) the holes must be strongly correlated.
 To be of negative sign an exchange interaction should be of inderect nature
( for example, via the oxygen ions which surround the magnetic ion).
But in this case
in a real system holes often prefer to hop not onto a magnetic ion but
rather to on a non-magnetic (oxygen). Two situations can arise with
increasing hole concentration. First: holes develop their own band,
and the localized moments of magnetic ions are not destroyed.
One then comes to a system with $s-d$-exchange,  which usually exhibites
some sort of magnetic order or Kondo-like effect. Second: the mixing
interaction between the orbitals of  magnetic and non-magnetic ions is
strong enough to destroy localized magnetic moments. This leads
to Mott delocalization, where the main role is played by the renomalization
of mixing interaction. In the $t-J$-model mixing is ``hidden" in {\em both}
hopping and exchange parameters, but it enters {\em in different powers}.
Thus,
for the description of these situations the $t-J$-model seems to be inadequate.

   The main conclusion of our work on the N\'eel AF state is that the
kinetic energy suppresses the magnetic order. This conclusion is of general
nature and is a familiar picture in the behaviour of various  transition
metal and actinide compounds.  Here, it manifests itself in decreasing
the critical concentration
$\delta _{c}(t)$ with increasing $t$. This  is not connected with any
type of frustration with increase of $\delta $ but rather it is due to
increasing degree
of delocalization of electrons. To get a feeling of it one can write down
the mean-field equation for magnetization for the simpler case of
ferromagnetic order ($J<0$) and rectangular density of states
($g_0(\epsilon) = 1/2D$ for $|\epsilon |\leq D$, and $g_0(\epsilon) = 0$
otherwise, $2D$ is the
bandwidth). Excluding the chemical potential with the help
of (45) one obtains
\begin{equation}
\frac{sinh \beta m (2Jg_0 + 1)}{sinh \beta m (2Jg_0 - 1)} = e^{\beta (1 -
\delta)D}.
\end{equation}
As seen from (61), the critical temperature
\begin{equation}
T_c^{(FM)} = \frac{D(1 - \delta)}{ln \frac{(2Jg_0 + 1)}{(2Jg_0 - 1)}}
\end{equation}
vanishes at the point $2Jg_0 = 1$. One recognizes the Stoner criterium
obtained, however, from the strong-coupling side (atomic limit
perturbation theory). Therefore, in the case of suffitiently large
bandwidth ferromagnetism does not arise. The conclusion is that only
the peak in the density of states (Figure 3)
which has a width $\delta D < J$ can provide
magnetism. In our case of N\'eel AF the spectrum $\propto cos k_x + cos k_y$
yields a van Hove singularity
and the region with  $\delta D < J$ exists  for any
hopping $t$. However with increasing $t$ the afm region in $(t-\delta )$-plane
and the value of moment itself becomes so small, that it must be destroyed
either by the gain in kinetic energy, or by fluctuations. The question of
the role of quantum fluctuations in the case of doped AF requires special
investigation. We believe, however, that this feature should survive in a more
adequate theory:
the total
number of states is constant and increasing
hopping decreases the
number of relevant states inside of the
peak. Thus, the magnetization
must be decreased.

\newpage
\subsubsection*{Acknowledgements}
One of us (I.S.) thanks T.M.Rice and G.Blatter for support, stimulating
discussions, and hospitality during a visit at the ETH Z\"urich.
The authors are grateful to J.K\"ubler for his hospitality,  the TH
Darmstadt and the SFB 252, for support.
Support of this work from the IFW Dresden is also very much acknowledged.
We thank B. Brandow, S.L.Drechsler, P.Entel, and A. M\"obius
for helpful discussions and for critical
comments concerning
the manuscript.

\newpage
\subsubsection*{References}
{[}1]~~Anderson, P.W., in Proc. Int. Enrico Fermi School of Phys. (1988);\\
Rice, T.M., Z. Phys. B {\bf 67} (1987) 141.\\
{[}2]~~Anderson, P.W., Science {\bf 235} (1987) 1196.\\
{[}3]~~Zhang, F.C., and T.M.~Rice, Phys. Rev. B {\bf 37} (1988) 3759.\\
{[}4]~~Ovchinnikov, S.G., and I.S.~Sandalov, Physica C {\bf 161} (1989) 607.\\
{[}5]~~Batyev, E.G., \v{Z}ETF {\bf 82} (1982) 1990 (Soviet Phys. --- JETP
{\bf 55} (1982) 1144).
---, \v{Z}ETF {\bf 84} (1983) 1517 (Soviet Phys. --- JETP {\bf 57}
(1983) 884).\\
{[}6]~~Wiegmann, P.B., Phys. Rev. Lett. {\bf 60} (1988) 821.\\
{[}7]~~Perelomov, A., {\em Generalized Coherent States and Their Applications},
Springer-Verlag, Berlin Heidelberg 1986.\\
{[}8]~~Coleman, P., {\em Moment Formation in Solids}, ed. W.J.L.~Bayers,
Plenum Press, New York 1984, p.~279.\\
{[}9]~~Read, N., D.M.~Newns, and A.C.~Hewson, ibid., p. 257.\\
{[}10]~Gutzwiller, M.C., Phys. Rev. Lett. {\bf 10} (1963) 159.\\
{[}11]~Rice, T.M., and K.~Ueda, Phys. Rev. Lett. {\bf 55} (1985) 997.
---, Phys. Rev. B {\bf 34} (1986) 6420.\\
{[}12]~Zhang, F.C., C.~Gros, T.M.~Rice, and H.~Shiba, Supercond. Sci. Technol.
{\bf 1} (1988) 36.\\
{[}13]~Zhang, Fuchun, {\em Superconducting Instability of Staggered Flux Phase
in the $t$-$J$-model}, Preprint 1989. \\
{[}14]~Lederer, P., D.~Poilblanc, and T.M.~Rice, {\em Correlated Electron
Motion, Flux-States and Superconductivity}, Preprint ETH-TH/89-32,
Z\"urich 1989. Phys.Rev.Lett. {\bf 63 } (1989) 1519.\\
{[}15]~Khveshchenko, D.V., and Y.I.~Kogan, Mod.~Phys.~Lett. B {\bf 4} (1990)
95.
JETP. Lett. {\bf 50} (1989) 137.\\
{[}16]~Kotliar, G., Phys. Rev. B {\bf 37} (1988) 3664.\\
{[}17]~Affleck, I., and J.B.~Marston, Phys. Rev. B {\bf 37} (1988) 3774.\\
{[}18]~ The alternative  is to describe the system from the very beginning in
$\vec{k}$-space, but to write 3-particle Fadeev equations for the vertices as
it has been suggested
by A.E.Ruckenstein, St.Schmitt-Rink in  Int.J.Mod.Phys. B {\bf 3} (1989) 1809.
However, this can be used only for non-zero doping and hopping. \\
{[}19]~Weller, W., Proc. Int. Sem. on High Temperature Superconductivity,
Dubna 1989, ed. by
 V.L.~Acsenov, N.N.~Bogolubov, and N.M.~Plakida,
World Scientific, Singapore 1989, p. 458.\\
{[}20]~noted by T.M.~Rice.\\
{[}21]~Barnes S.E., Physica B {\bf 165 \& 166} (1990) 987.\\
{[}22]~Barnes, S.E., J.Phys.F{\bf 6} (1976) 115. 1375. {\em ibid.}{\bf 7}
(1977) 2637; Adv. Phys. {\bf 30}, (1980) 801.\\
{[}23]~Z.Zou, P.W.Anderson, Phys.Rev. B{\bf 37} (1988) 627.\\
{[}24]~Zaitsev, R.O., \v{Z}ETF {\bf 70} (1976) 1100 (Soviet Phys. --- JETP
{\bf 43} (1976) 574).\\
{[}25]~The speculations (13)-(15) can be applied to this case directly.\\
{[}26]~Recently, it has been described once more in details by Yu.A. Izyumov
and B.M.
     Letfulov in, J. Phys.: Condens. Matter {\bf 2} (1990) 8905.\\
{[}27]~Gor'kov, L.P., \v{Z}ETF {\bf 34} (1958) 735 (Soviet Phys. --- JETP
{\bf 7} (1958) 505).\\
{[}28]~Nambu, Y., Phys. Rev. {\bf 112} (1958) 812.\\
{[}29]~X.G.Wen, Frank Wilczek, A.Zee, Phys. Rev. B{\bf 39} (1989) 11413.\\
{[}30]~D.Coffey, K.S.Bedell, and S.A.Trugman, Phys. Rev. B{\bf 42} (1990)
6509.\\
{[}31]~N.Garsia, A.Levanyuk and P.Serena,Preprint (Univ.Autonoma de Madrid,
28049).\\

\newpage

%this is Figure 1

\begin{figure}

\caption{ The individual atoms of
one elementary cell are labeled by $\alpha=(1,2,3,4)$.
Sites 1 and 3 belong to the sublattice A (with
spin projection $\uparrow$), sites 2 and 4 to the sublattice B ($\downarrow$).
}
\end{figure}
\newpage

%this is Figure 2

% GNUPLOT: LaTeX picture
\begin{figure}
\setlength{\unitlength}{0.240900pt}
\ifx\plotpoint\undefined\newsavebox{\plotpoint}\fi
\sbox{\plotpoint}{\rule[-0.175pt]{0.350pt}{0.350pt}}%
%
\caption{ Dependence of the N\'eel magnetization $m$ on the
concentration of holes $\delta$ at temperature $T=0$
for various hopping parameters $t$ (given
in units of $J$; those lines where no $t$-value is indicated refer
to $t=2.$ and $t=3.$, respectively).}
\end{figure}

\newpage

%this is Figure 3

\begin{figure}
\begin{center}
% GNUPLOT: LaTeX picture
\setlength{\unitlength}{0.240900pt}
\ifx\plotpoint\undefined\newsavebox{\plotpoint}\fi
\sbox{\plotpoint}{\rule[-0.175pt]{0.350pt}{0.350pt}}%
%
\end{center}
\caption{Density of states, $g(\epsilon)$ (in arbitrary units),
for the high-moment state (upper
graph) and for the unstable low-moment state (lower graph).
The related parameter values are $t=1.$, $\delta=0.25$, and $T=0$
in both cases.
Thick lines refer to the occupied part of the bands.
The energy, $\epsilon$, is given in units of $J$.}
\end{figure}

\newpage

%this is Figure 4

\begin{figure}
\begin{center}
% GNUPLOT: LaTeX picture
\setlength{\unitlength}{0.240900pt}
\ifx\plotpoint\undefined\newsavebox{\plotpoint}\fi
\sbox{\plotpoint}{\rule[-0.175pt]{0.350pt}{0.350pt}}%

\end{center}
\caption{Zero-temperature phase diagram.
Antiferromagnetism exists in the region 2.
The curve separating regions 2 and 3 is the branch-curve of the
magnetic solutions, see Figure 2.
At low values of $t$, the magnetization changes here from a finite
value to zero.
At $1.5 < t < 2.$, the character of the transition changes to second
order type.
On the curve between regions 1 and 2, the transition is of second
order, too.}
\end{figure}

\end{document}